# AsicBoost -
# A Speedup for Bitcoin Mining

Dr. Timo Hanke

March 31, 2016 (rev. 5)

**Abstract.** *AsicBoost* is a method to speed up Bitcoin mining by a factor of approximately 20%. The performance gain is achieved through a high-level optimization of the Bitcoin mining algorithm which allows for drastic reduction in gate count on the mining chip. *AsicBoost* is applicable to all types of mining hardware and chip designs. This paper presents the idea behind the method and describes the information flow in implementations of *AsicBoost*.

# 1 Introduction

*AsicBoost* is a method to speed up Bitcoin mining by a factor of approximately 20%. *AsicBoost* is an *algorithmic* optimization and therefore applicable to all types of mining hardware.

The *AsicBoost* method is based on a new way to process work items inside and outside of the Bitcoin mining ASIC. It involves a new design of the SHA 256 hash-engines (inside the ASIC) and an additional pre-processing step as part of the mining software (outside the ASIC). The result is a performance improvement of up to 20% achieved through a reduction of gate count on the silicon. The purpose of this paper is to present the idea behind the method and to describe the information flow in implementations of *AsicBoost*.

The hash-engine design required for *AsicBoost* is compatible with design philosophies such as pipelined ("unrolled") cores and non-pipelined ("rolled") cores. The performance gains can be achieved on top of all low-level optimizations regarding timing, pipelining, path balancing, custom cell and full-custom designs, etc.

Through gate count reduction on the silicon *AsicBoost* improves two essential Bitcoin mining cost metrics simultaneously and by a similar factor: the energy consumption (Joule per Gh) and the system cost ($ per Gh/s). With the system cost being proportional to the capital expenses of a Bitcoin mine and the energy consumption being proportional to its operating expenses, *AsicBoost* reduces the total cost per bitcoin mined by approximately 20%. For the Bitcoin mines of the future *AsicBoost* will make all the difference between a profitable and an unprofitable mine.

A thorough analysis of all algorithmic Bitcoin mining optimizations before *AsicBoost* has been done in [5]. The fact that [5] estimates the combined performance gain of all optimizations it

http://asicboost.com/
asicboost@gmail.com

describes at 7% demonstrates the significance of *AsicBoost*. Moreover, some of the previous optimizations are compatible with *AsicBoost* in a way that their performance gains add up.

The *AsicBoost* method was invented by Timo Hanke in collaboration with Sergio Demian Lerner and is patent-pending.

# 2 Preliminaries on Bitcoin Mining

## 2.1 SHA 256

SHA 256 is a cryptographic hash function from the SHA 2 family defined by NIST [1]. The SHA 256 digest of a message is calculated based on dividing the message into *chunks* of 64 bytes each, and processing the chunks consecutively through a state machine. The final state after having processed all chunks yields the SHA digest of the original message.

During processing, each chunk is run through a *message expander* function which produces a corresponding *message schedule* of 64 words[1]. The message schedule is then fed into a *compressor* function which changes its internal state in the process of consuming the message schedule in 64 rounds, one word at a time. The compressor's state after having processed all message schedules derived from all chunks of the original message turns into the final hash digest of the original message.

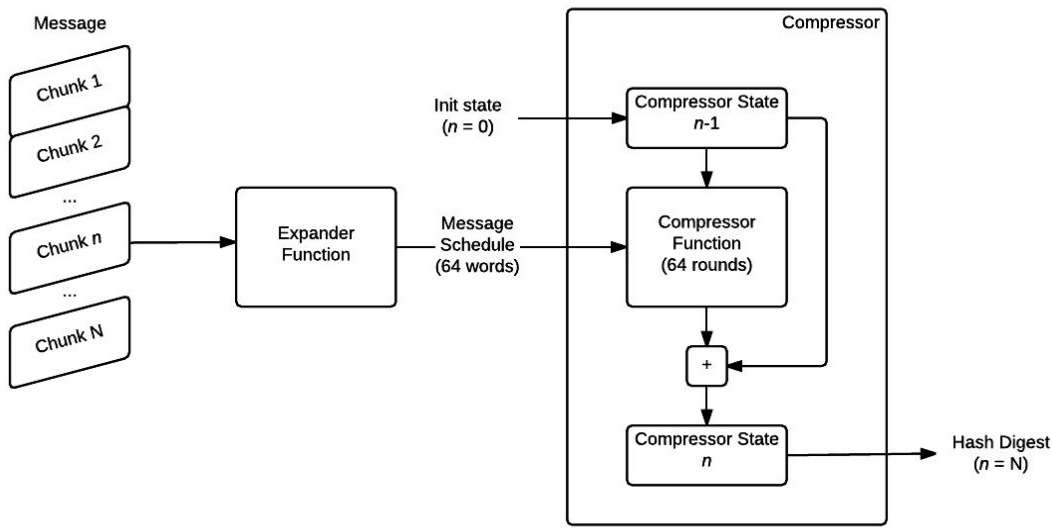

Fig. 1. SHA 256

---

[1] A *word* is 4 bytes.

http://asicboost.com/
asicboost@gmail.com

## 2.2 The Bitcoin block header

Bitcoin has been introduced by S. Nakamoto in [2]. A good source for Bitcoin's protocol specification and technical descriptions is the Bitcoin wiki [3]. Of specific relevance to this paper is the Block hashing algorithm described in [4].

In Bitcoin mining, the message to be hashed is the *block header*. A Bitcoin block header is 80 bytes long and is divided into two chunks as follows, the second chunk being padded to a length of 64 bytes:

| Chunk 1 | | | | Chunk 2 | | | |
|---|---|---|---|---|---|---|---|
| Block header | | | | | | | Padding |
| Block header candidate | | | | | | Nonce | |
| Version | Previous hash | Merkle root | | Time stamp | Bits (difficulty) | | |
| | | Head | Tail | | | | |
| 4 bytes | 32 bytes | 28 bytes | 4 bytes | 4 bytes | 4 bytes | 4 bytes | 48 bytes |
| | | | Message[2] | | | | |

Fig. 2. The Bitcoin Block Header

## 2.3 Bitcoin's mining function

Bitcoin's mining function processes block headers with a double SHA, and the resulting information flow for both chunks through the double SHA is shown in Fig. 3 below.

---

[2] The first 12 bytes of Chunk 2 are called *Message* for the purpose of this document.

http://asicboost.com/
asicboost@gmail.com

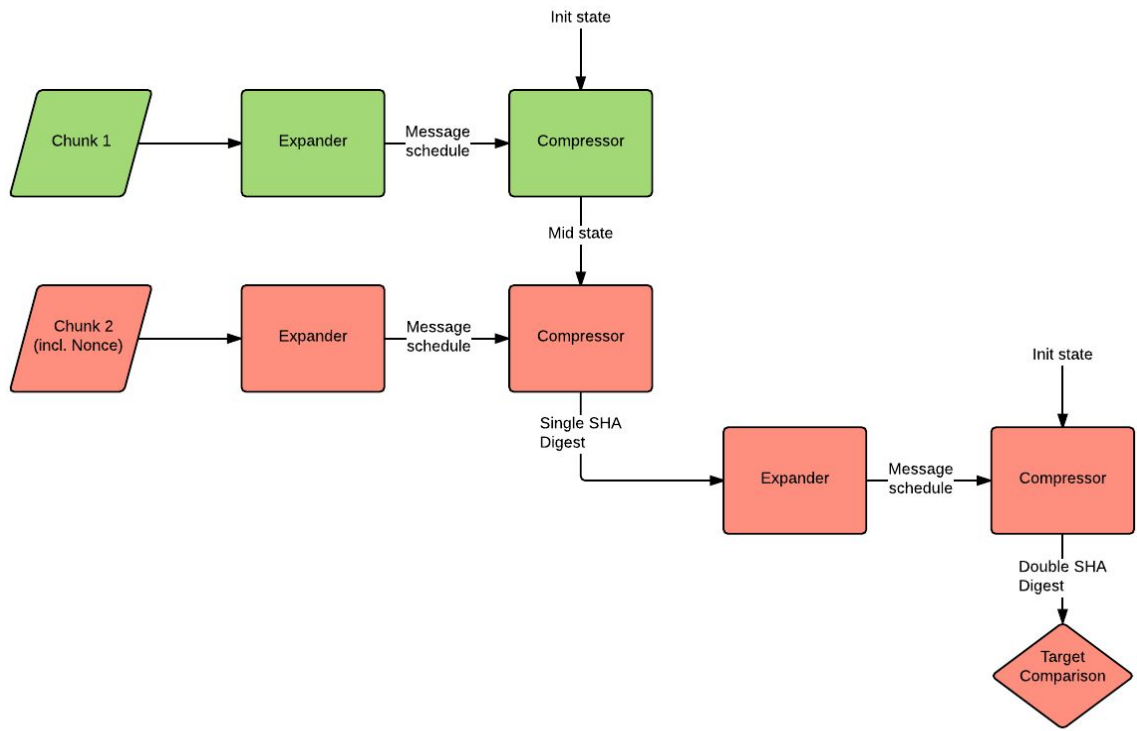

Fig 3. Bitcoin's Double SHA

The resulting double SHA digest is compared against the target range based on the current network difficulty, and if a match is found then the block header (including the *Nonce* that produced the match) is broadcast to the Bitcoin network.

## 2.4 Scanning the Nonce range

The Bitcoin mining process selects a *block header candidate*, loops over the whole Nonce range and filters out the Nonces that produce a double SHA digest matching the target. When the Nonce range is exhausted a new block header candidate is selected.

The part of the computation that is repeated in the loop over the Nonce range is depicted in red in Fig. 3. This part depends on the Nonce. The green part of Fig. 3 is not repeated in the loop as it depends only on the block header candidate, not on the Nonce.

The loop over the Nonce range starts out with this information, together called a *Work item*:
- Mid state
- Message

The Mid state is the output of the green compressor function. The Message is the part of Chunk 2 that depends on the block header candidate. The Message excludes the Nonce (which is selected inside the loop) and the padding (which is constant).

In light of this, the following diagram shows the core part of the computation involved in Bitcoin mining, or the *Bitcoin mining loop*:

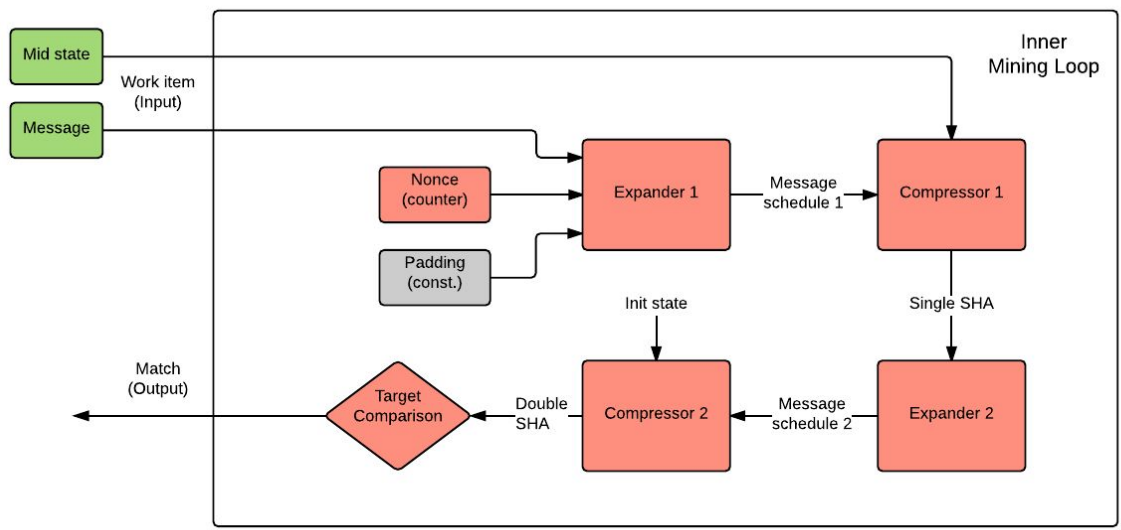

Fig. 4. Bitcoin's Mining Loop

The components shown in red lie in the *inner loop* and update at high frequency. The components shown in green lie in the *outer loop* and update less frequently.
All Bitcoin mining ASICs process only the inner loop internally. The work items are pre-computed and passed to the ASIC from outside.

## 2.5 Building work items

Each work item is constructed from a new block header candidate which in turn is constructed from a new *Merkle root*. Since the Merkle root spans both chunks of the block header (see Fig. 2), updating the Merkle root affects both components of the work item: Midstate (depending on chunk 1) *and* Message (depending on chunk 2). Therefore, it is usually not the case that different work items share either of the two components. This is the reason why the information in *Message Schedule 1* of Fig. 4 cannot be re-used across the processing of multiple work items.

# 3 AsicBoost

## 3.1 Colliding work items

*AsicBoost* achieves its performance gain by highly re-using Message Schedule 1 across multiple work items. In order to do so, *AsicBoost* generates many block header candidates that all share a common Message part. We speak of block header candidates as *colliding* if they collide in the Message part, or, in other words, if they differ from each other only in chunk 1.

The work items derived from colliding block header candidates all share a common Message component and differ only in the Mid state component. We speak of work items as *colliding* if they collide in the Message component.

## 3.2 The AsicBoost Loop

A set of colliding block headers allows *AsicBoost* to swap the inner and outer mining loops as shown in Fig. 5 below. The Message is constant throughout the processing of the entire set of colliding block headers. *AsicBoost's* inner loop iterates over the Mid states of all colliding block headers. The Nonce is now updated in the outer loop rather than the inner loop. Message schedule 1 is updated less frequently and its information is re-used across multiple Mid states.

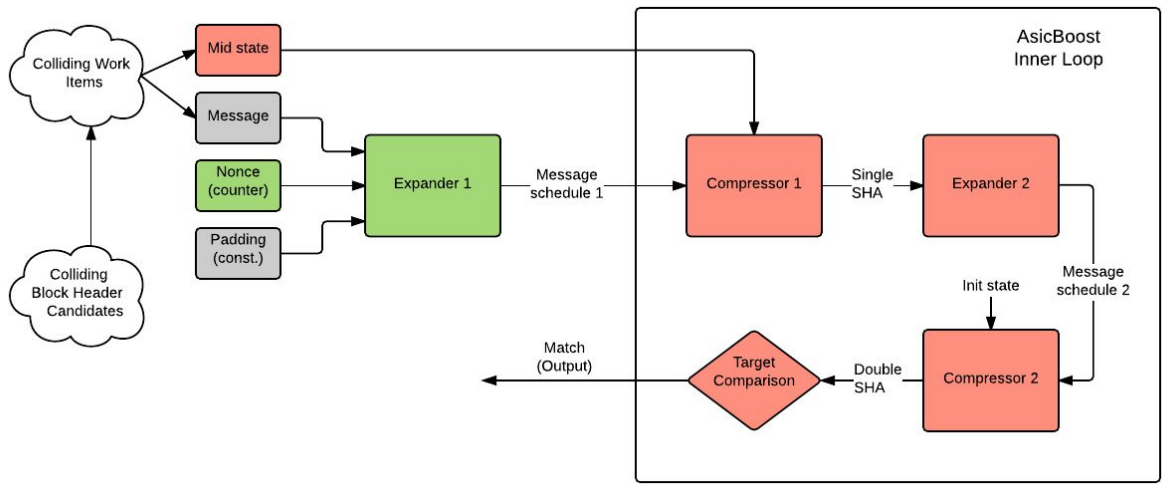

Fig. 5. AsicBoost's Mining Loop

## 3.3 The AsicBoost Gain

The AsicBoost method eliminates the Expander 1 function from the inner loop, leaving only Compressor 1, Expander 2 and Compressor 2 to be processed at high frequency. Since each of the four functions have similar complexity, up to one quarter of the total computational work can be saved by this approach.

The exact gain depends on the size of the set of colliding work items. Suppose the Expander 1 takes $x$ percent of the computational work of the four functions combined. Having $n$ colliding work items then gives a total gain of $x\,(n-1)/n$ percent. Assuming for simplicity that Expander 1 is one quarter of the four functions combined, then we get:

| Number of colliding work items ($n$) | 1 | 2 | 4 | 5 | 8 | 16 |
|---|---|---|---|---|---|---|
| Gain in percent | 0 | 12.5 | 18.75 | 20 | 21.9 | 23.4 |

Fig. 6. AsicBoost Gain Percentage

For comparison, [5] estimates the combined performance gain from all previously known optimizations to be 7%.

The implementation of *AsicBoost* in a mining chip is discussed in section 3.5 below. Some forms of ASIC implementation have to be designed for a specific value of n. It should be noted that *AsicBoost* can also be implemented in software, e.g. on GPUs. In fact, the availability of random access memory makes the implementation very easy and straight-forward without incurring much overhead.

## 3.4 Building colliding block header candidates

### 3.4.1 Merkle root collisions

There are several ways to produce block header candidates that differ only in chunk 1. One way is to calculate many merkle roots at random and filter them based on their last 4 bytes until sufficiently many are found that collide in their last 4 bytes.

There are a number of trade-offs to be considered when filtering for collisions. The optimal process depends on the targeted value for $n$ and the number and speed of the hashing cores to be served. We will not discuss the optimal solution in this paper, but further information can be provided by the author on request.

There are several methods to calculate merkle tree roots that are more efficient than changing the coinbase field. One method is based on permuting the order of transactions inside the block, or, in other words, permuting the leafs of the merkle tree. We will not discuss the optimal solution in this paper, but further information can be provided by the author on request.

### 3.4.2 Alternatives

Another, more efficient way to produce colliding block headers is by using bits inside Chunk 1 but outside the merkle root that are free for the miner to choose. This does not require finding any merkle root collisions. Instead, one merkle root is chosen and fixed. The free bits are then updated in a loop and the respective Mid states are computed from Chunk 1. Each Mid state obtained in this way gives a new colliding work item, so that this method is extremely efficient.

While the Bitcoin protocol currently does not define any bits outside the merkle root as "free", there are hardfork proposals that would make a number of bits at the beginning of the *Previous hash* component free. These bits are currently always zero and the proposal would re-define them as an extra nonce.

## 3.5 AsicBoost chip design

A Bitcoin mining ASIC consists of several hashing cores that work in parallel, either simultaneously on the same work item or each on on their own work item. The block diagram of a traditional hashing core looks identical to what is shown in Fig. 4 as the Inner Mining Loop. For a description of a traditional mining chip also see [6].

### 3.5.1 The Multi-Core Design

The easiest way to adopt *AsicBoost* is to share an Expander 1 block among two or more hashing cores. An example for two cores is shown in Fig. 7 below, where two cores sharing an Expander 1 turn into a single unit called a *Duo-core*. In this design, the Mid state generation is kept outside of the ASIC and the work distribution on the chip is equivalent to the work distribution on a traditional chip.

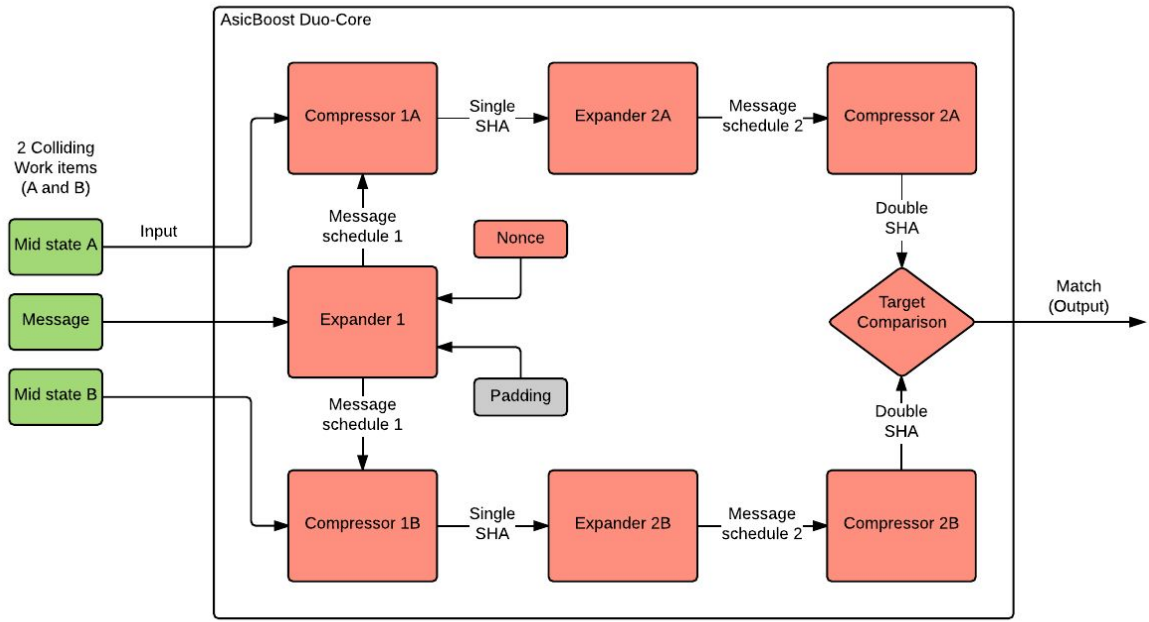

Fig. 7. AsicBoost Duo-core

The example of Fig. 7 can be generalized from a Duo-core to a *Multi-Core* in which the Expander 1 output is shared among $n$ cores. The expected performance gain can be seen in Fig. 6 above and is about 12.5% in the case of Duo-cores.

There are a number of trade-offs to be considered when designing Multi-Cores. The optimal implementation and optimal value for $n$ will depend on many low-level factors of the original hashing core design. The details are beyond the scope of this paper, but further information can be provided by the author on request.

### 3.5.2 The Low-Toggle Design

In another implementation each core has its own Expander 1 logic, but this logic is toggling at lower frequency. An example is shown in Fig. 8 below. Multiple Mid states are fed to the core and are processed completely before the Nonce updates. When the Nonce has updated the same set of Mid states can be processed again. The Expander 1 logic toggles only when the Nonce or the Message updates, saving most of the energy that would otherwise be consumed by Expander 1.

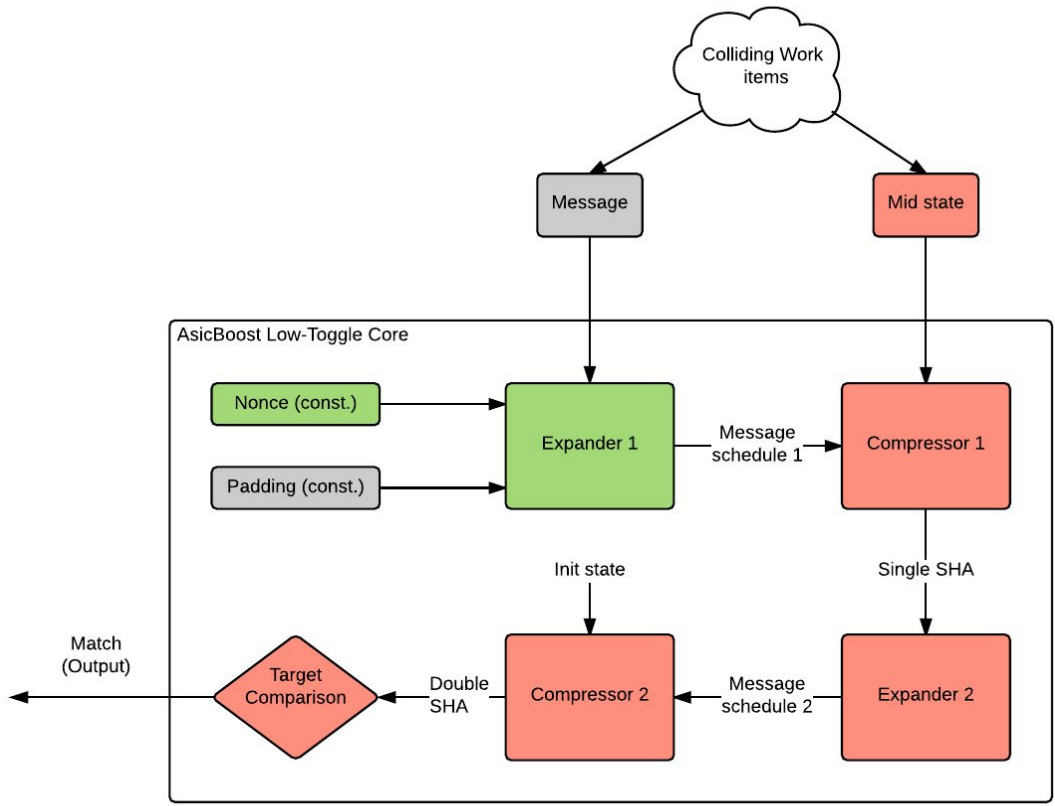

Fig. 8. AsicBoost Low-Toggle Core

There are a number of trade-offs to be considered regarding the distribution of multiple Mid states at high speed to the cores. The optimal implementation will depend on many low-level factors of the original hashing core design. The details are beyond the scope of this paper, but further information can be provided by the author on request.

# 4 References

1. *SHA-2*, Wikipedia, https://en.wikipedia.org/wiki/SHA-2.
2. *Bitcoin: A Peer-to-Peer Electronic Cash System*, Satoshi Nakamoto, White paper, 2008, https://en.bitcoin.it/wiki/Bitcoin_whitepaper.
3. Bitcoin wiki, https://en.bitcoin.it/wiki.
4. *Block hashing algorithm*, Bitcoin wiki, https://en.bitcoin.it/wiki/Block_hashing_algorithm.
5. *Optimising the SHA256 Hashing Algorithm for Faster and More Efficient Bitcoin Mining*, R. Naik, MSc thesis, 2013.
6. *Goldstrike 1: CoinTerra's First-Generation Cryptocurrency Mining Processor for Bitcoin*, IEEE Micro 35 (2), J. Barkatullah and T. Hanke, 2015.

Contact:
Dr. Timo Hanke
http://asicboost.com/
asicboost@gmail.com